\documentclass[a4paper]{article}

\usepackage[english]{babel}
\usepackage[utf8]{inputenc}
\usepackage{amsmath}
\usepackage{graphicx}
\usepackage[colorinlistoftodos]{todonotes}
\usepackage{chapterbib}
\usepackage{float}
\usepackage{authblk}
\usepackage{url}
\usepackage{hyperref}

\usepackage{color}
\usepackage{ulem}

\title{Blockchain Inefficiency in the Bitcoin Peers Network}
\author[1,2]{Giuseppe Pappalardo\footnote{Correspondence author: g.pappalardo@ucl.ac.uk}}
\author[1,3,4]{T. Di Matteo}
\author[2]{Guido Caldarelli}
\author[1,3]{Tomaso Aste}
\affil[1]{Department of Computer Science, UCL, London, UK}
\affil[2]{IMT School for Advanced Studies, Lucca, IT}
\affil[3]{UCL Centre for Blockchain Technologies, UCL, London, UK}
\affil[4]{Department of Mathematics King's College London, London, UK}

\begin{document}
\maketitle

\begin{abstract}
We investigate Bitcoin network  monitoring the dynamics of blocks and transactions. 
We unveil that 43\% of the transactions are still not included in the Blockchain after 1h from the first time they were seen in the network and  20\% of the transactions are still not included in the Blockchain after 30 days, revealing therefore great inefficiency in the Bitcoin system.
However, we observe that most of these `forgotten' transactions have low values and in terms of transferred value the system is less inefficient with 93\% of the transactions value being included into the Blockchain within 3h.
The fact that a sizeable fraction of transactions is not processed timely casts serious  doubts on the usability of the Bitcoin Blockchain for reliable time-stamping purposes and calls for a debate about the right systems of incentives which a peer-to-peer unintermediated system should introduce to promote efficient transaction recording.

\end{abstract}

\section{Introduction}

Behind Bitcoin  \cite{nakamoto2009bitcoin}, the most popular cryptographic currency, there are users distributed all over the world who, in a voluntary way or for profit, participate in a network where transactions are announced, verified and eventually inserted into blocks of a massively replicated ledger known as Blockchain \cite{Antonopoulos:2014:MBU:2695500}.
The Blockchain is a distributed database which keeps track of all payments made by using the Bitcoin currency.
Payments are called ``transactions'' and involve one or more input Bitcoin addresses which are sending some funds to one or more ``output'' addresses.
Despite the  success of this new approach, which has seen bitcoin becoming the most important cryptocurrency with capitalization exceeding fifteen billions US dollars, the system is far from being optimal. 
At the time of writing of this paper (September 2016) there were around $6,000$ peers  participating to the bitcoin network. Blocks contained typically  between $1$ and $1.7$ thousand transactions, counting for about 100-170 transactions per minute. 
These transactions mobilised a capital of about 152 Bitcoins per minute (about $91,787$ USD/minute at  September '16 exchange rate). Bitcoin blockchain is based on a  mechanism of a peer validation for the different blocks by the majority of computational power  \cite{nakamoto2009bitcoin} (the ``proof of work'') which can be considered one of the most important innovations introduced by bitcoin. 
Indeed, it solves several issues related to trust and machine synchronisation that are otherwise hard to manage in a distributed system operating between untrustful peers \cite{nakamoto2009bitcoin,Antonopoulos:2014:MBU:2695500,Weber2016,bitcoininnovations,ADTIEEE2017}.
The operation to include transactions inside blocks and their validation is performed by special nodes called ``miners'' which participate to the solution, by brute force, of a cryptographic task consisting in finding a hash number associated with the block content which is smaller than a given target number (the above mentioned ``proof of work''). 
Miners are incentivized to do this operation with a reward in newly issued bitcoins given to the first who find and successfully broadcast the valid hash. 

In this paper we analyse some of the properties of this self-organised system that could be, in principle, improved.
In particular, we monitor transactions and blocks exchanged among peers during a period of slightly more than seven days (from 04/05/2016 and 11/05/2016). We then measure the time that takes to these transactions to be correctly recorded into valid blocks and become part of the blockchain.  
We observe that  most transactions are recorded in the Blockchain after a few blocks with 58\% of transaction recorded within one hour. However we also observe a sizeable amount of transactions that take a much longer time to be inserted with 20\% of transactions left still unrecorded after 30 days.
We argue that, the way the bitcoin consensus mechanism is designed, although has proven its effectiveness, it also introduces intrinsic inefficiencies. 
In particular, miners are incentivised to verify transactions but there are no mechanisms that ensures that all transactions are actually processed.
Miners can freely chose to include or not to include transactions in the blocks; although they have no particular advantage or interest to exclude transactions they also do not have specific incentives to verify that all transactions broadcasted in the network are included into the blocks.

\subsection{Blockchain} 
In the bitcoin network, whenever a transaction is made, it is then announced by broadcasting it to a series of peers that send the information to their own contacts which propagate the announce further.
Peers also validate transactions which are gathered into blocks which are cryptographically sealed and inserted (every $10$ minutes approximately) into the Blockchain after a validation from the community. 
At the core of the Blockchain technology there is this consensus mechanism where peers agree on the order of the transactions with a process called ``proof of work" similar to voting by computation power majority and that consists in solving a cryptographic problem where it is given an hash associated to the block of transactions and to the  hash of the previous block (that are in this way sequentially ``chained'').
In theory every peer could participate to this consensus mechanism but, in the years, this activity has become typical of a specialised part of the community called ``miners''. Miners get newly emitted Bitcoin in reward for this activity.
A block has inside the hash of the last valid block and the record of the most recent transactions observed by the miner and not included yet in the Blockchain.
The miner will  try to seal it cryptographically with a hash produced from the block itself and a random part. 
If the hash number is by chances smaller than a threshold imposed by the proof-of-work then it is considered ``valid'' and it can start to be broadcasted to the network.
When a peer receives a new block, it should verify if the block is valid.
In order to do that, it has to verify whether the hash of the block fulfils the proof-of-work requirements. 
After that, the peer has to also verify the digital signatures and the formatting of each transaction  included inside the block. 
If the whole block and all the transactions are verified, it accepts the new block as valid and starts propagating it through the network (and if the peer is a miner, also it will start to discover the next block on top of it). If the block is not valid, or at least one transaction inside the block is invalid, the block will be discarded.
The Blockchain is the chain of blocks built one on top of the other in chronological sequence uniquely associated with a sequence of hash numbers.
Miners get their gain only from the cryptographic sealing of new blocks with a valid hash number; therefore, they have no incentives to make the system efficient by carefully checking {if} all transactions are included in the blocks.  

\subsection{Bitcoin communication protocol}
All Bitcoin clients are connected to each other in a peer to peer network. 
This means that there are no central servers or authorities.
Each node individually decides how to contribute to the network by choosing which service to provide. 
For example, by relaying transactions,  by storing a copy of the Blockchain or by using their own computational power for mining.
A node wanting to join to the network for the first time needs to connect to some special peers called ``seeds''. Such seeds nodes provide a list of peers known to this ``seed'' node. This list does not depend on geographic location of clients; all the clients included are chosen randomly and can contain up to one thousand nodes.
After retrieving the peers list, a node chooses peers until it reaches its default max number of connections (usually from $8$ to $126$ established connections, but the number of connections may vary according to the configuration of the Bitcoin client used and to the network setting of the client itself).
Once connected to the network, a node can send and receive messages (such as blocks, transactions and new peers joined on the network) from all the other connected nodes. All these messages have to follow the rules (that may have different customisation) settled up by the Bitcoin Protocol \cite{bitcoinprotocol}, which consists of a set of messages used by clients to enable communication among peers.

\section{Related Work}
In the last few years there has been some interest in the study of the Bitcoin network with two notable contributions from Decker \cite{Decker2013Information} and Miller \cite{coinscope}. 
There are also online services, such as Blockchain.info \cite{blockchain.info} which allows users to explore blocks and transaction.
Instead, the platform Bitnodes \cite{bitnodes} provides snapshots of all reachable peers on the network and some statistics related to the type of the client (i.e. protocol version used, last block stored and ip-geolocalization). 
Since all the data are provided as a list of online clients, it is impossible to reconstruct how the peers are connected to each other or how information propagate among them.
The approach used to discover peers on the Bitcoin network is to send recursively ``getaddr'' message to each reachable node in order to get back part of their known nodes list.  In 2015 Andrew Miller published a Bitcoin network investigation called Coinscope \cite{coinscope} which used this approach in order to discover clients,  introducing also an algorithm, named ``AddressProbe'' which was able to track how peers were connected. At that time, before the release of Bitcoin Core 0.10.1 \cite{guessingbitcoin}, discovering connections was possible because each client kept updated the timestamp of a peer in a ``mempool'' after each data exchange. 
Every time a client replied back to a peers list, it  also sent their updated timestamps.
The mechanism for updating the timestamp was the following: if a node exchanged some messages with a peer, it keept its own timestamp on the updated database. If instead a node discovered some new nodes through another peer, it applied a two hours penalties on the timestamp before storing the address into its own peer database.
Through this mechanism it was possible to guess the connections of a peer just retrieving several time the known peers list and sorting all the records in chronological order \cite{DBLP:journals/corr/BiryukovKP14}.
However, Biryukov \cite{DBLP:journals/corr/BiryukovKP14,guessingbitcoin}  showed that reconstructing  peers network could be used to attack  Bitcoin Core clients. To avoid the possibility of such attack, the software was modified such that each client does not necessarily update  the timestamp of a connected client when they send or receive data. 
After last update on the client we noticed that, for an active connection, the timestamp is updated only when the connection drops or after 24 hours (in the case the connection is still alive). All the other cases are still as described in \cite{coinscope}.

Data propagation rate in the Bitcoin network was studied by Decker et al. in \cite{Decker2013Information} where {, by establishing connections with each node, they measured  the time that blocks or transactions was received take to propagate into the network.}

In this paper, we follow this methodology to identify the appearance of blocks and transactions in the network and we measure the propagation dynamics in the network and the time they take to be included inside the Blockchain.

\section{Methods}
To monitor the Bitcoin network we setup a customised client able to recursively establish a connection with each reachable node, requesting its known peers list and  trying to connect to them and retrieve their list.
To accomplish this goal, our client did not need to implement the whole protocol, but only a reduced set of messages:
\begin{itemize}
\item{{\em getaddr, addr}} - {The ``getaddr'' messages is used to request a list of known peers from a node. The node will issue an ``addr'' message as response containing up to one thousand known nodes. The ``addr'' messages are also sent automatically to each connected node  when the client establish a connection with a new node.}
\item{{\em inv}} - The ``Inventory'' message is sent by a client when it discovers new blocks or transactions in order to spread them on the network. In the same {\em ``inv''} message it is possible to have blocks and transactions together.
\end{itemize}
The {\em ``addr''} messages are required in order to connect to all reachable peers and to new discovered peers once they join the network.
Once connected, we stored all the inventory messages received in the form:
{\it{timestamp, address, hashcode}};
where timestamp is a 64bit integer representing the time and date when the {\em ``inv''} message was received, address is the ip address of the nodes (which can belong to ipv4, ipv6 or tor networks) and the hashcode is the hashing string corresponding to a block or to a transaction.
We are establishing only one connection to each peer and we do not make any {\em ``getdata''} request in order to not add load on the network. This approach has the drawback that each peer can close the connection at any time without sending any alert. When this happens, we suddenly try to establish a new connection. The information shared with the other peers during the time in which the connection was down is lost.

Data exchanged by peers consist of coordinating signals (i.e. announcing new blocks or transactions) and data messages (blocks, addresses and transactions).
Data were collected by joining  the network as a normal node and trying to establish a connection within each peer address discovered and waiting for {\em ``inv''} messages for both, blocks and transactions.
The client for collecting the data was written in Go programming language \cite{gopl}.

\section{Data}
We {monitored} the Bitcoin network activity during the period from Wed, 04 May 2016 01:20:45 GMT to Wed, 11 May 2016 18:44:58 GMT. 
During this  period slightly longer than 7 days, we observed over twelve thousands unique peers, $8,969$ belonging to ipv4 network, $3,332$ belonging to ipv6 network and $124$ belonging to Tor network, with an average of $5-7$ thousands client connected at the same time. This amount of peers is consistent with the amount reported by Bitnodes \cite{bitnodes}.
Surprisingly, we received from the peers more than $126$ thousands different blocks some of them valid but ``old'', where the oldest of them were included into the Blockchain  more than $3$ years earlier.
Instead, the number of blocks mined during the listening period was $1,209$ valid blocks (from block height $410,119$ to $411,327$).
Overall we collected 592GB of data with the most part regarding transactions {\em ``inv''} messages (589 GB) while the remaining  related to blocks {\em ``inv''} messages.
During the investigation time we observed $12,424$ different nodes. Each node can relay blocks and transactions. We received both of them (blocks and transactions) from $11,537$ nodes.
There are also nodes which relay only blocks or only transactions, indeed we received blocks from $11,537$ nodes and transactions from $12,168$ nodes.
We decided to classify blocks and transactions as follow:
\begin{itemize}
	\item{\bf Blocks}
	\subitem{\bf{Mined During Listening Block (MDLB)}} - This set identifies all the blocks which were included on the Blockchain during the listening period and propagated by the peers before the next block was discovered. There are 1,209 blocks discovered by 530 source nodes and spread through 11,179 destination nodes. The maximum number of blocks discovered by a single node during the listening time was 86. These are the only blocks analysed.
	\subitem{\bf{Echo Blocks (EB)}} - This set identifies all the blocks already included in the Blockchain and propagated in delay. There were  $406,457$ echo blocks, propagated from $6,938$ nodes. 	
	\subitem{\bf{Fork Block (FB)}} - This set identifies all the blocks  not included in the Blockchain even if they had a valid hash. There were $34$ fork Blocks of this kind.
	\subitem{\bf{Invalid Block (IB)}} - This set identifies all the blocks not included in the Blockchain, but propagated by the peers despite having a hash above the proof-of-work threshold (not valid). There were $51,103$ Invalid Blocks transmitted by $23$ nodes. 
	\item{\bf Transactions}
	\subitem{\bf{Blockchain Transactions (BT)}} - Valid transaction, included in the Blockchain, observed and propagated through the network before the block in which they get eventually included is discovered.  We received $1,744,899$ Blockchain Transaction, from which $1,725,508$ were included in a block during the listening time and $19,391$ were included in a block after the listening time. 
We discarded transactions observed for the first time during the mining of the first and the last block and also those which were received after they were included into a block. We also did not analyse transactions with a set locktime (about five thousands). The final subset of transactions we analyzed count $64,994$ which were generated by $2,518$ nodes.
	\subitem{\bf{Echo Transaction (ET)}} - Valid Transaction, already included in a Block but still propagated in delay. We received $12,425$ echo transactions that were not analysed.
	\subitem{\bf{Invalid Transaction (IT)}} - Transaction not valid for some reasons. We received 62,889 Invalid transactions that were not analyzed.
\end{itemize}

\section{Results and Discussion}
Firstly we measured the time needed to mine the valid block hashes (MDLB). 
We found that the minimum time is about $2$ minutes, while the maximum time is $77$ minutes;  the medium time for discovering a block is about $9$ minutes and the $50\%$ percentile is about $6$ minutes.
In table \ref{timing} we make a comparison between the block receiving time in a peer and the time actually recorded in the Blockchain.
In the table, the minimum (maximum) time is the time required for a block to be discovered and it is negative due to a fork event. Medium Time is the average discovery time considering all blocks listened. Their variance and percentile are also reported.
Tables \ref{protocols} and \ref{useragent} report respectively the protocols and the Bitcoin client used by the nodes of the network.
\begin{table}
\centering
\begin{tabular}{|l|r|r|}
\hline
                & Listening Time & Blockchain  Time\\
\hline             
Minimum Time    & -5.48 s    & -558 s\\
Maximum Time    & 4650.09 s  & 4642 s\\
Medium Time    & 550.05 s   & 550.05 s\\
Variance        & 550.11 s   & 550.30 s\\
Percentile 50\% & 383.25 s   & 384 s\\
\hline
\end{tabular}
\caption{\label{tab:widgets} This table shows some statistics related to ``mined during listening'' blocks set, comparing timestamp reported on each block  with the time reported inside the Blockchain. The time on the Blockchain can be wrong since a miner could vary the timestamp if the nonce is not sufficient to produce a valid hash number. The minimum time is negative due to a Fork event. }
\label{timing}
\end{table}

\begin{table}[htbp]
\begin{center}
	\caption{Bitcoin Protocol version used by nodes}
	\begin{tabular}{|l|r|}
		\hline
		Protocol & number of clients \\ \hline
		70012 & 6655 \\ \hline
		70002 & 3013 \\ \hline
		N/A & 771 \\ \hline
		7000 & 1153 \\ \hline
		70013 & 88 \\ \hline
		70010 &78 \\ \hline
		80001 & 71 \\ \hline
		70011& 68 \\ \hline
		80000& 24 \\ \hline
		99999&4 \\ \hline
		50400&2 \\ \hline
		70014&2 \\ \hline
		60000&1 \\ \hline
		70003&1 \\ \hline
		60002&1 \\ \hline
		80002&1 \\ \hline
	\end{tabular}
	\label{protocols}
	\end{center}
\end{table}
\begin{table}[htbp]
\begin{center}
	\caption{Bitcoin Client Software used by 20 or more nodes}
	\begin{tabular}{|l|l|r|}
		\hline 
		Bitcoin Software and Version & Number of clients \\ \hline
		 Classic:0.12.0 & 2969 \\ \hline
		Satoshi:0.12.1 & 1790 \\ \hline
		 Satoshi:0.12.0 & 1691 \\ \hline
		Satoshi:0.11.2 & 1323 \\ \hline
		 N/A & 771 \\ \hline
		 Satoshi:0.11.0 & 368 \\ \hline
		 Satoshi:0.10.2 & 233 \\ \hline
		 Satoshi:0.11.1 & 226 \\ \hline
		Classic:0.11.2 & 184 \\ \hline
		 Satoshi:0.12.99& 144 \\ \hline
		 Satoshi:0.11.2(bitcore)& 142 \\ \hline
		 Satoshi:0.9.3& 122 \\ \hline
		 Satoshi:0.10.0& 116 \\ \hline
		 BTCC:0.12.1& 93 \\ \hline
		 Satoshi:0.8.6& 72 \\ \hline
		 Satoshi:0.9.1& 71 \\ \hline
		BitcoinUnlimited:0.12.0(EB16; AD4)& 70 \\ \hline
		 Satoshi:0.10.1& 67 \\ \hline
		 Satoshi:0.9.2.1 & 56 \\ \hline
		 Satoshi:0.8.5 & 42 \\ \hline
		 Bitcoin XT:0.11.0D & 29 \\ \hline
		Bitcoin XT:0.11.0 & 26 \\ \hline
		Bitcoin XT:0.11.0E(Linux; x86\_64)/\&22', & 22 \\ \hline
		 Satoshi:0.8.1 & 20 \\ \hline
	\end{tabular}
	\label{useragent}
	\end{center}
\end{table}

\begin{figure}[h]
\begin{center}
	\includegraphics[width=\textwidth]{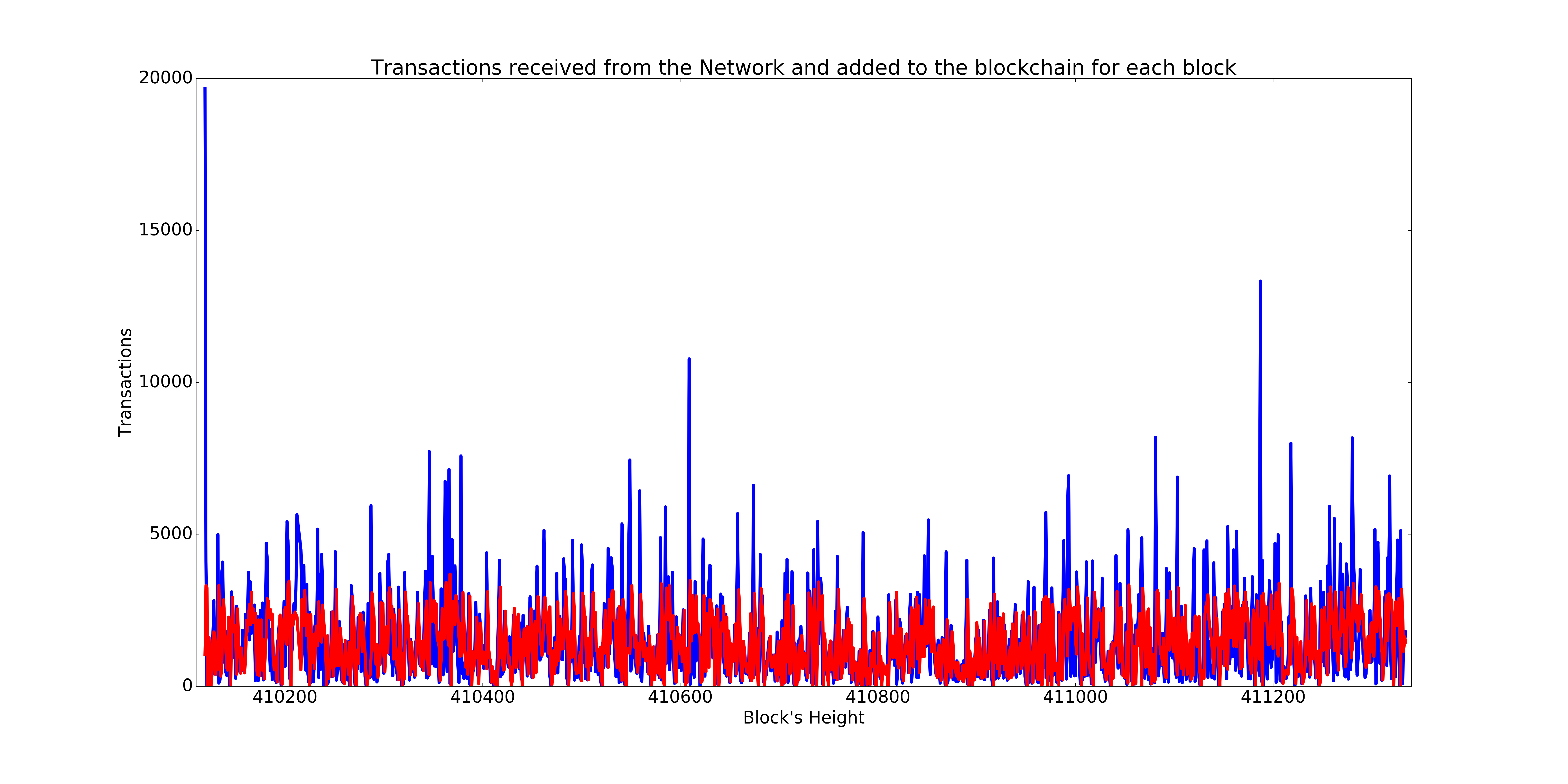}
	\caption{This figure compares the number of Transactions observed during block mining (in blue) and  the number of transactions included in the Blockchain during the same block mining time (in red). }
	\label{transazioniperblocco}
		\end{center}
\end{figure}

We also investigated both the transaction dynamics and the block dynamics on the Bitcoin network.
In Fig. \ref{transazioniperblocco} we compare the number of transactions observed during the listening period and the number of transactions included in the blocks during the same period. We observed that there are different dynamics. 
We show in the following that during this process some transactions are not included into blocks for a considerably long time. 

\subsection{Blocks}
\begin{figure}[bthp]
\begin{center}
\includegraphics[width=0.49\textwidth]{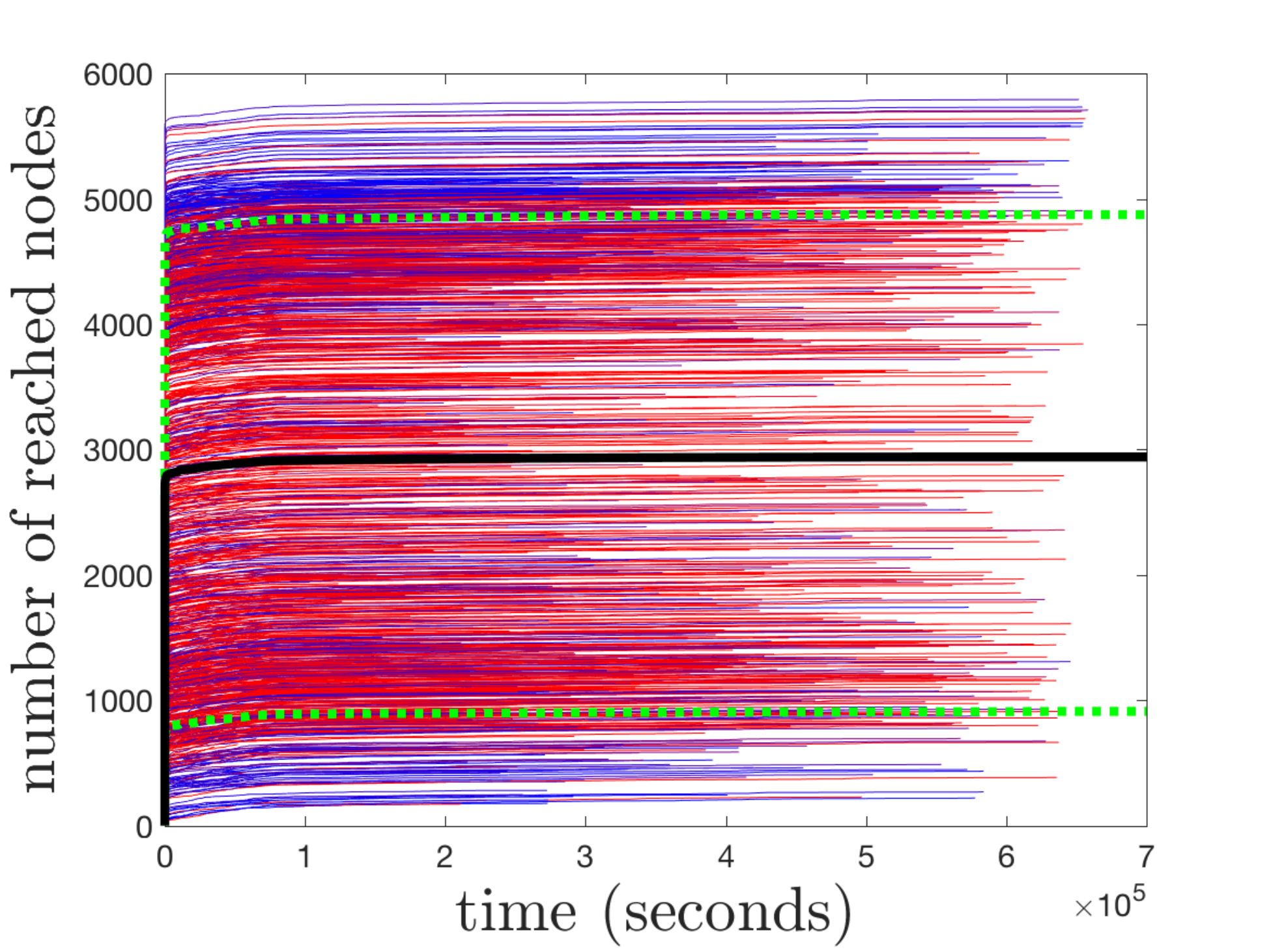} 
\includegraphics[width=0.49\textwidth]{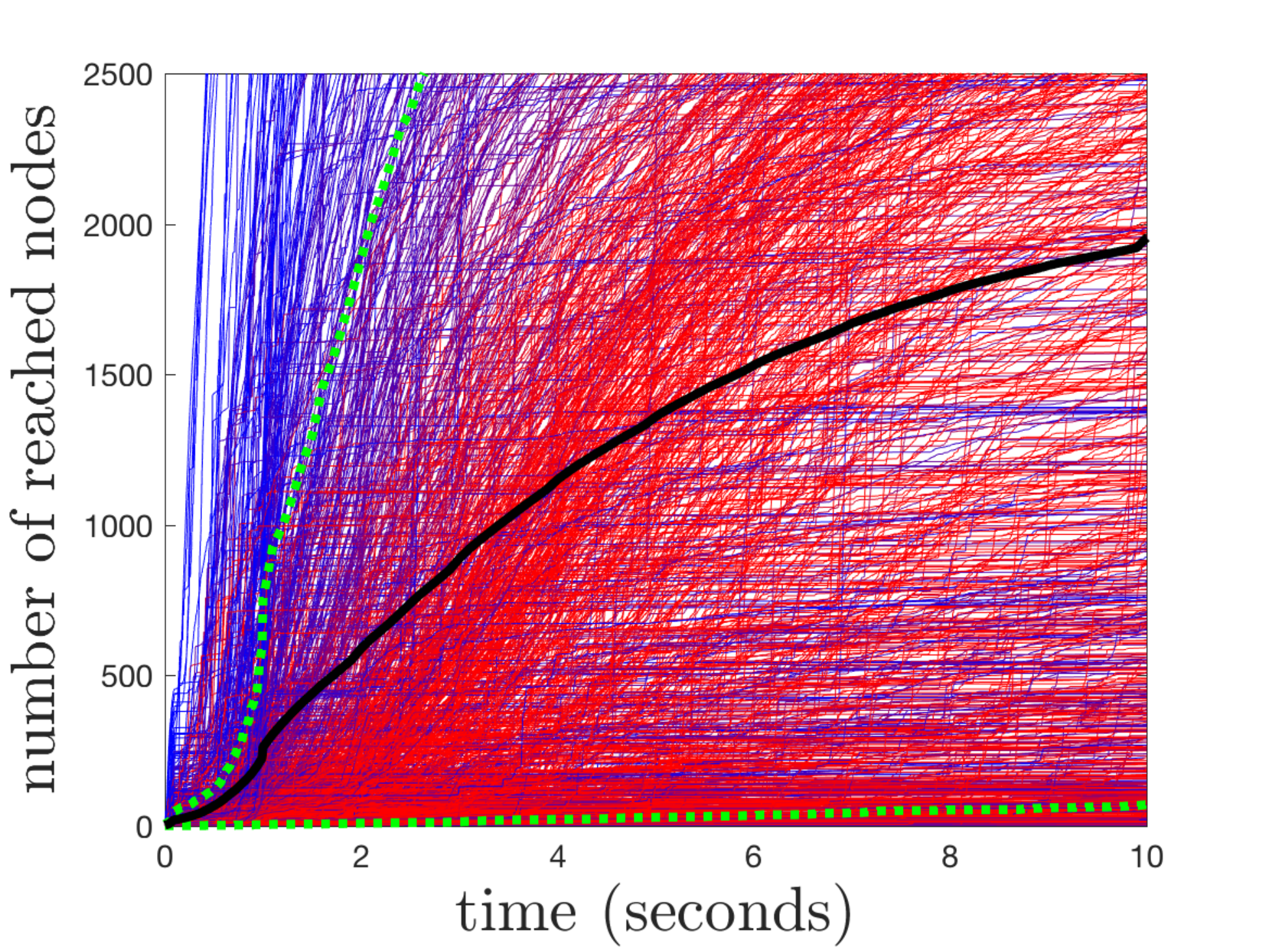}
	\end{center}
\caption{
Number of nodes reached by a new valid block before a following block is discovered (left).
The colours are associated with the size of the block: the size increases passing from blue to red.
The black line is the average of all the observations and the two dashed green lines are respectively the $10\%$ (lower) and $90\%$ (upper) percentiles. The right plot is a detail of the initial propagation within the first $10$ seconds.
}
\label{cumulative}
\end{figure}

In Fig. \ref{cumulative} we report, {for each block,} the number of peers reached  vs. the time lapse since the block was first observation in the Bitcoin network. 
Each trace of the plot terminates when a new valid block is discovered, for this reason the curves  have different durations {around 10 min in average}.   
We observe that different blocks reach a different number of nodes which depends on the propagation dynamics, on the number of nodes present during the propagation and on the number of active connections established by our client during the block propagation time.  
We observe that the typical propagation consists in a fast initial increase during the first second of propagation when about $10\%$ of nodes are reached. 
In this initial phase, the average propagation law (black line in the plot) is consistent with an exponential growth which is what is expected for a diffusion process over a random network \cite{EgaliatrianNetPaper}. 
Eventually, the process slows down reaching about  $60\%$ of nodes within the first 10~s of propagation and then it slows even further with the rest of the nodes  reached after several minutes.

We observe that  blocks first announced by some nodes propagate  consistently faster (or slower) than others.
We also observe that, at the extremes, blocks first propagated by the fastest node reach $50\%$ of the peers in $2.3$~s whereas blocks first propagated by the slowest node reach $50\%$ of the peers in more than $1,800$~s. 
Client type is another factor in the propagation process; however, the limited statistics does not allow us to determine more precisely this effect. 

\begin{figure}
\begin{center}
\includegraphics[width=\textwidth]{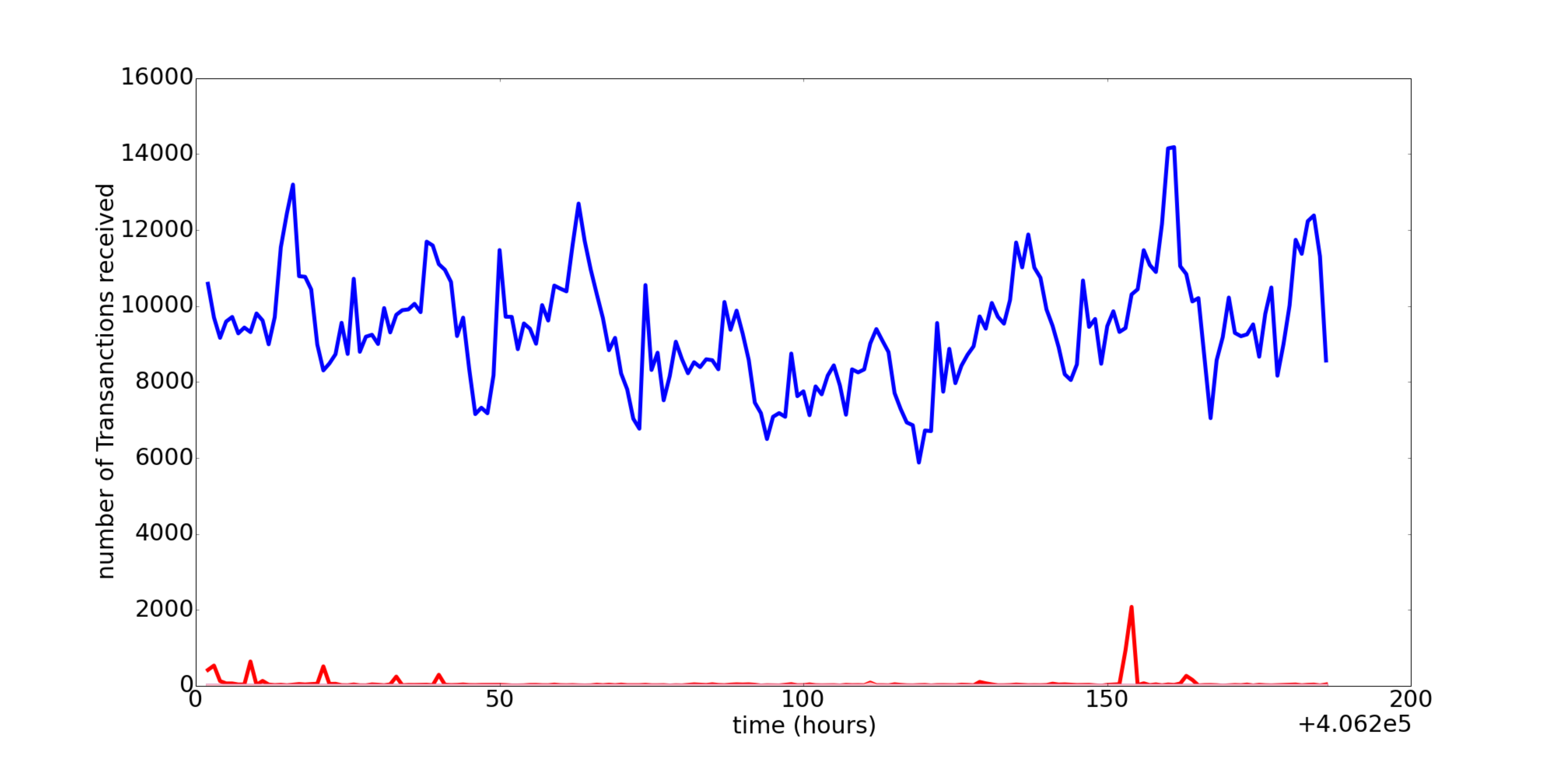}
\caption{Number of Transactions per hour received during the listening time.
The Blue line represent the transactions included in the Blockchain during or after the listening time (BT). 
The red line represent  the invalid transactions (IT).
}
\label{transazioni}
	\end{center}
\end{figure}

\begin{figure}
\begin{center}
\includegraphics[width=0.5\linewidth]{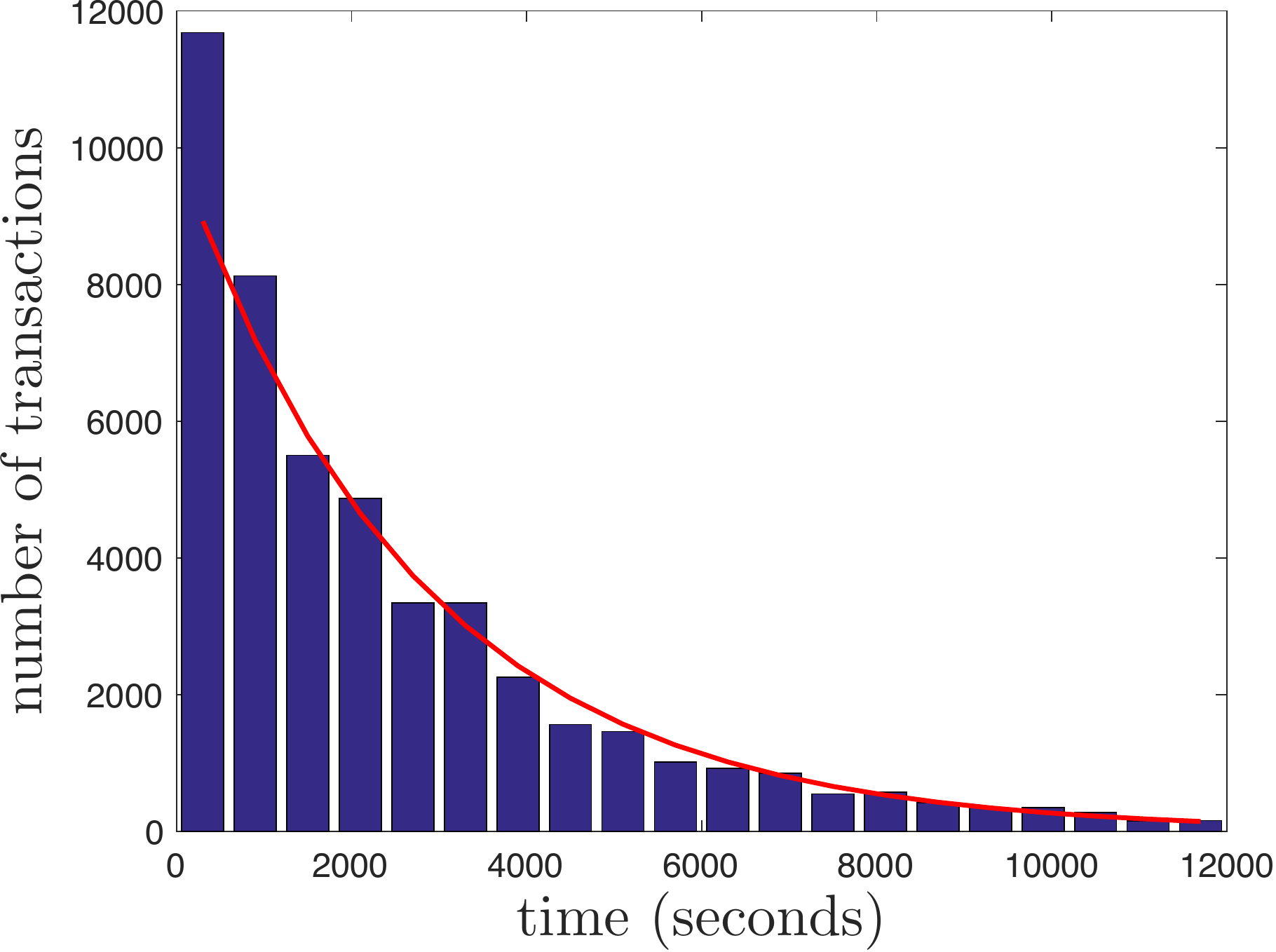}\includegraphics[width=0.5\linewidth]{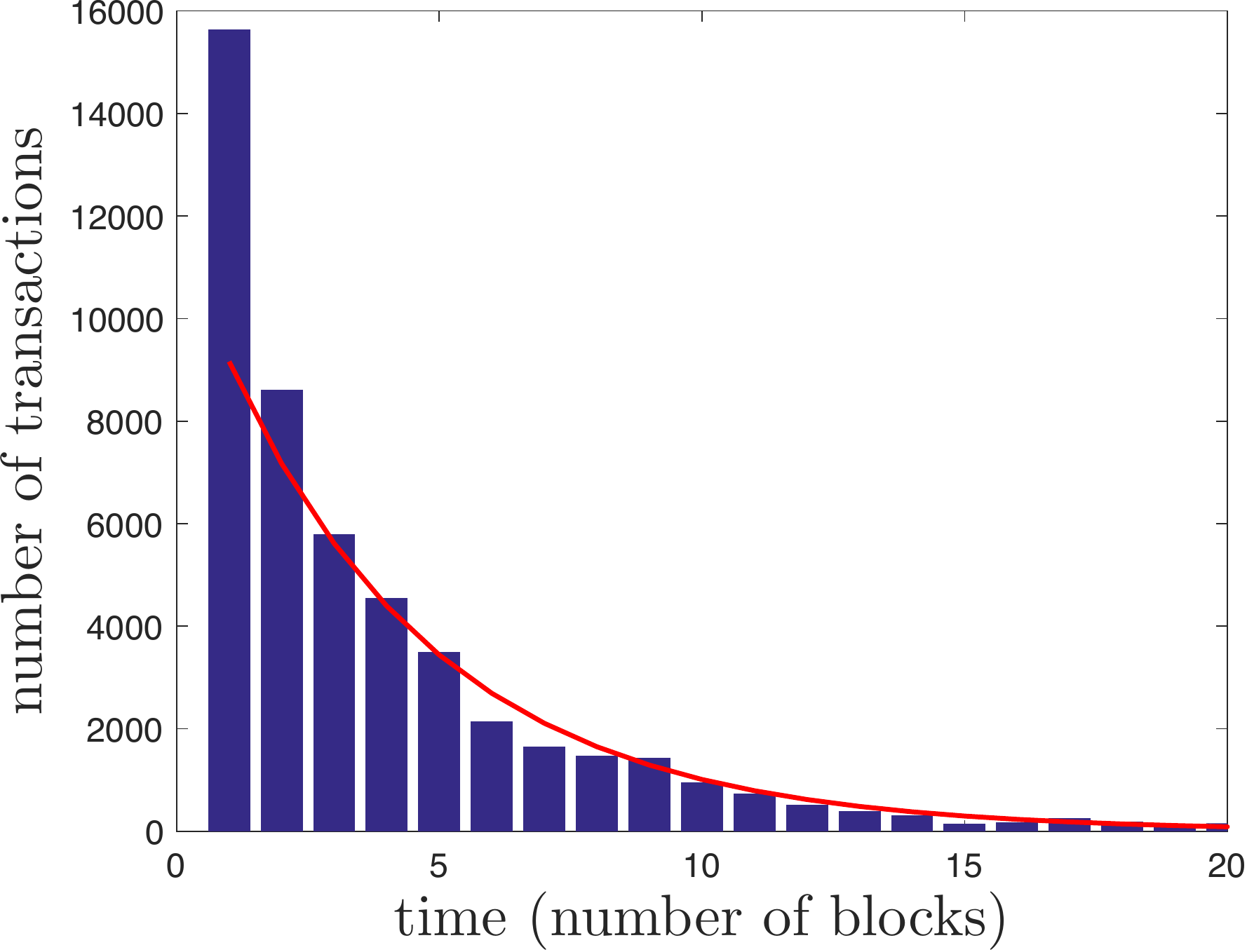}
\caption{Distribution of time intervals between the first time a transaction is observed in the network and the time when it is included into a valid block.
The left plot reports time in seconds and the right plot reports time in  number of blocks (approx 10 min each).
The red line are best fits with exponential decay law.}
\label{Tempotransazioni}
	\end{center}
\end{figure}

\begin{figure}
\begin{center}
\includegraphics[width=0.7\textwidth]{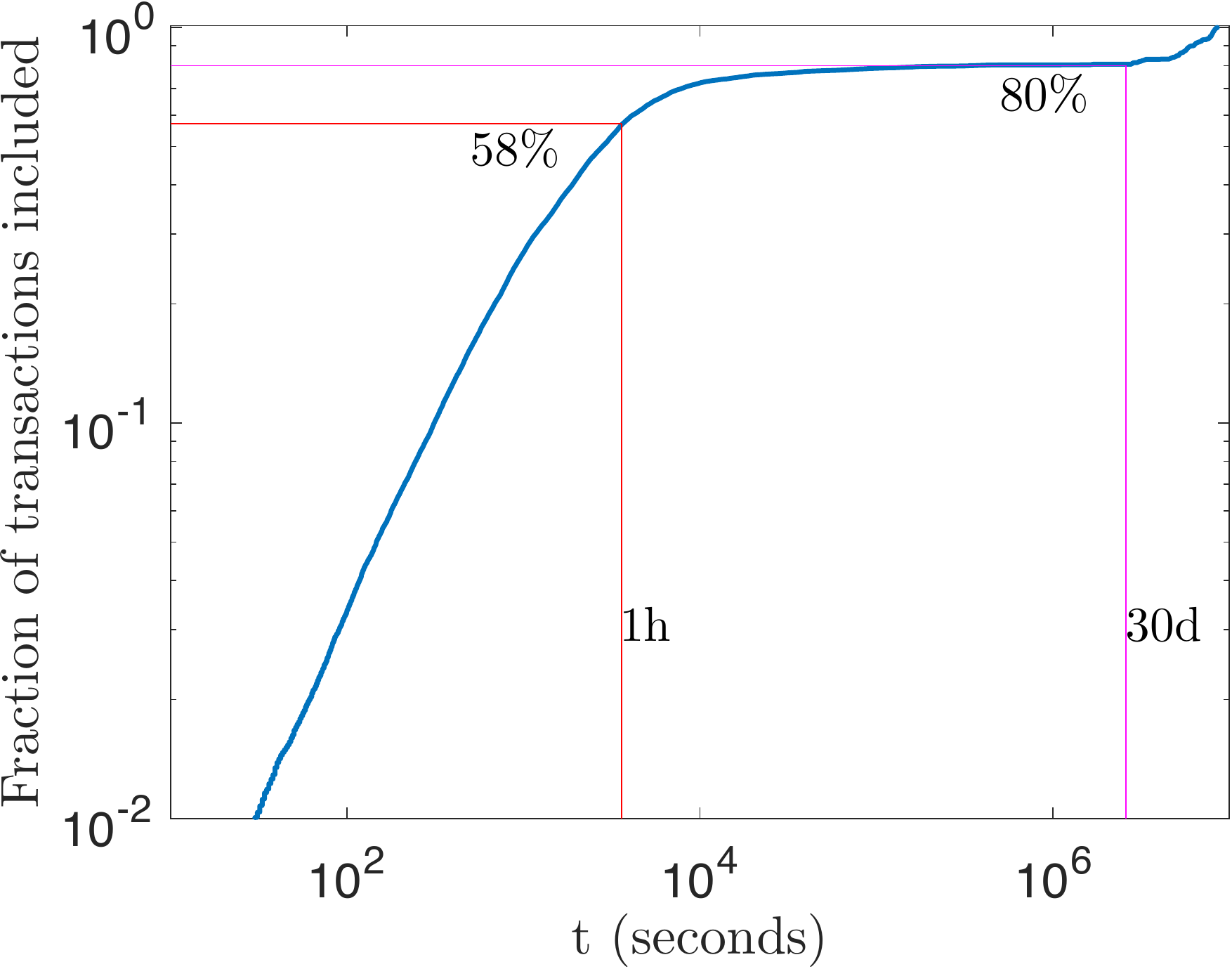}
\caption{
Fraction of transactions included in the Blockchain after a given amount of time (seconds, x-axis) from first observation in the network.
The two vertical lines mark 1h and 30 days.
}
\label{cumulativeInclusion}
	\end{center}
\end{figure}

\begin{figure}
\begin{center}
\includegraphics[width=0.7\textwidth]{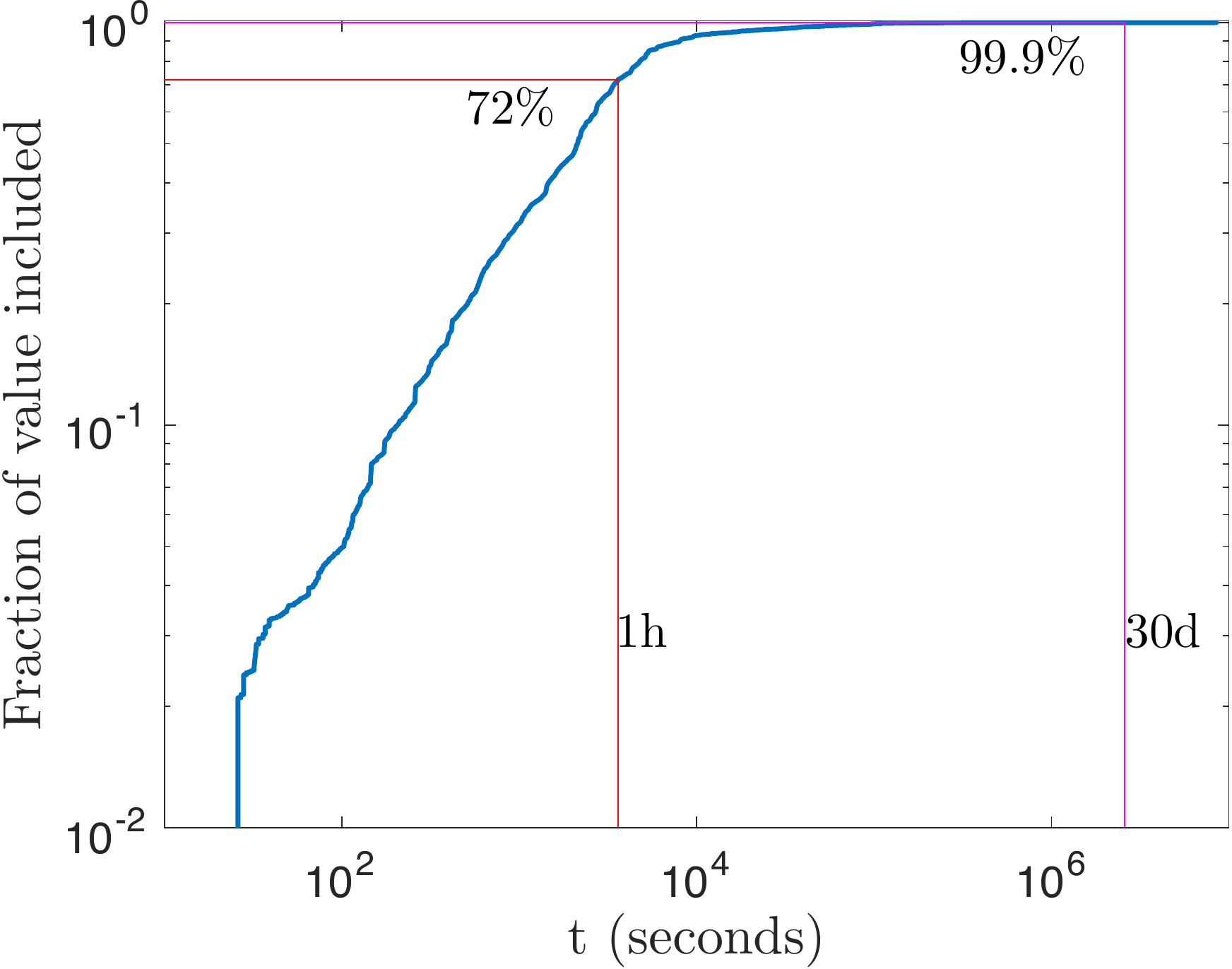}
\caption{
Fraction of transferred value included in the Blockchain after a given amount of time (seconds, x-axis) from first observation in the network.
}
\label{cumulativeInclusionValue}
	\end{center}
\end{figure}

\subsection{Transactions}
We recorded the transaction id received from clients together with the client address, and the time. 
Fig. \ref{transazioni} shows the received transactions rate per hour for IT set (in red) and BT plus ET set (in blue). 
During the listening time we received $1,820,212$ Transactions; $1,722,696$ of them were included in the Blockchain in the period until Sat, 09 Jul 2016 10:52:38 GMT. The Blockchain contains other $1,266$ transactions that have been produced during the monitoring period but were not observed by us in the network. An amount of $1,208$ of them corresponds to the zero-th transaction of each block (which is not broadcasted), whereas we could not observe the remaining $58$ ones. 

\begin{figure}
\begin{center}
\includegraphics[width=0.7\textwidth]{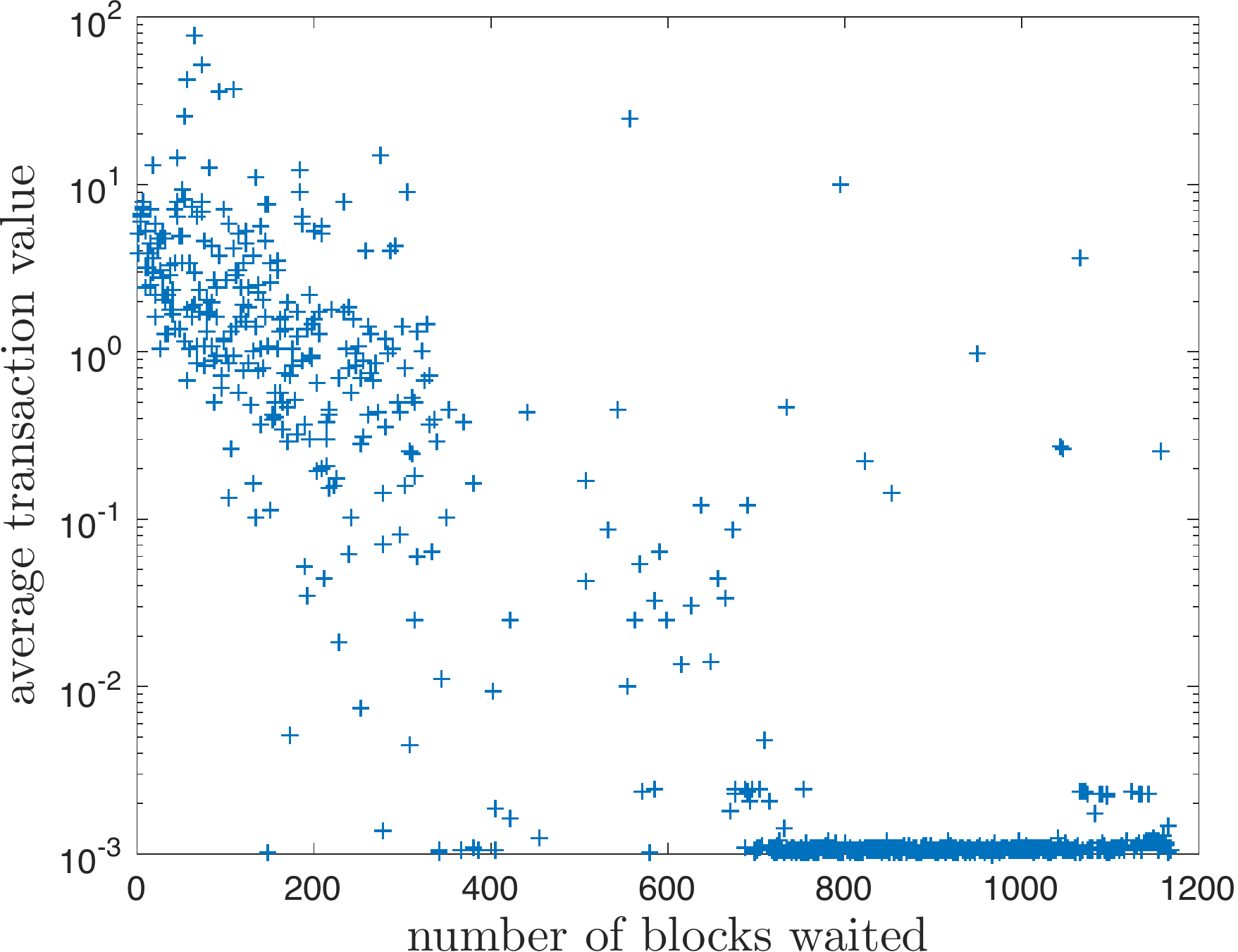}
\caption{
Average value of the transaction (in bitcoins) vs. waiting time in blocks numbers.
}
\label{BlocksValue}
	\end{center}
\end{figure}

\begin{figure}
\begin{center}
\includegraphics[width=0.7\textwidth]{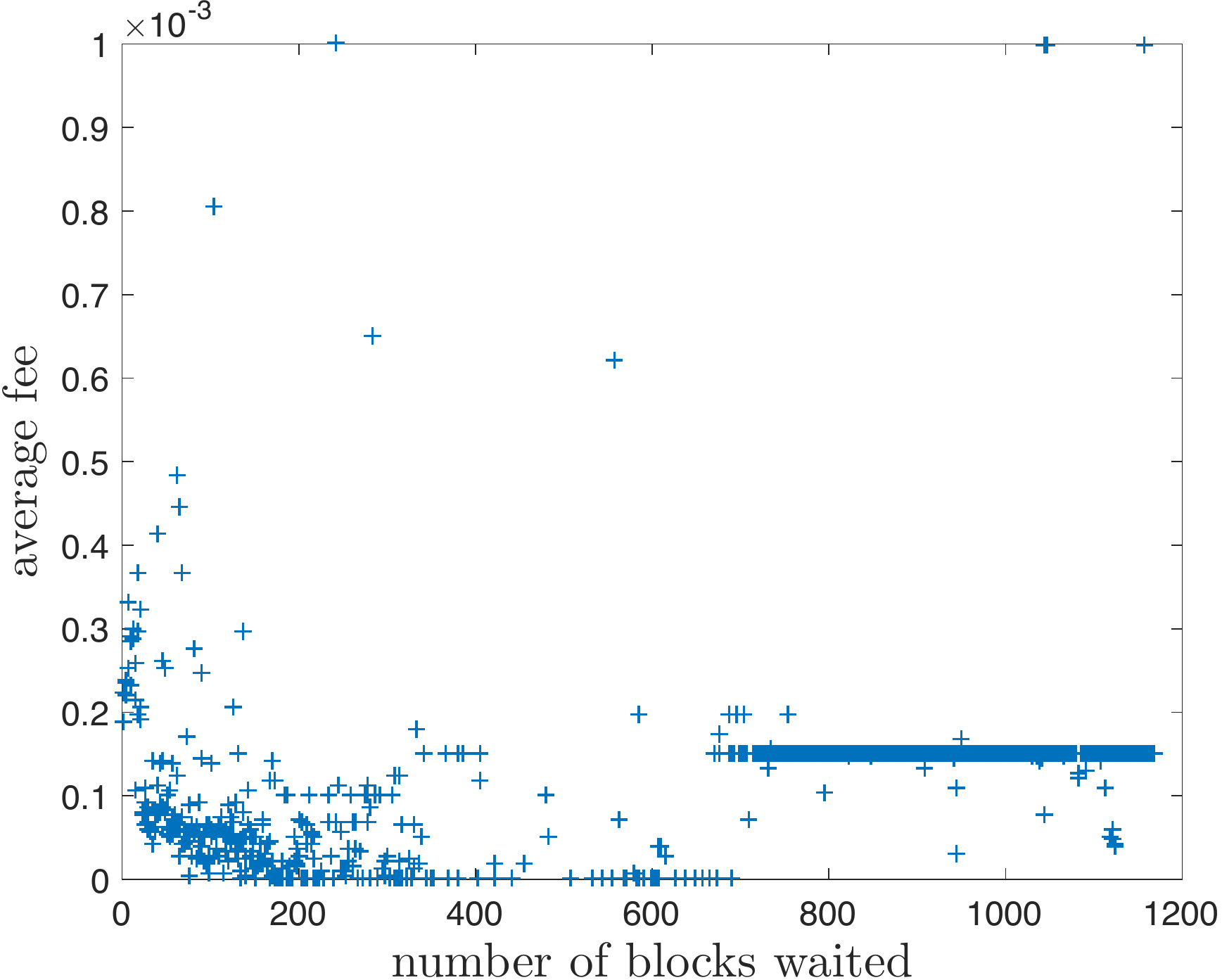}
\caption{
Average fee vs. waiting time in blocks numbers.
}
\label{BlocksFee}
	\end{center}
\end{figure}

We computed the interval of time between the first time a transaction is seen in the network and the time when it is included in the Blockchain: Fig. \ref{Tempotransazioni} shows the distribution of such intervals of time measured in seconds (left) and in number of blocks (right). We observe a decreasing behaviour  which is compared with the best-fitting exponential decay $\sim e^{(-t/\Delta)}$ reported in the figure with the red line.
 The coefficient $\Delta$ is the characteristic time and it was measured to be respectively $2,800$ s and $4.1$ blocks.
 However, we can observe from Fig. \ref{Tempotransazioni} that  the empirical time distribution does not follow precisely an exponential decay; instead, it tends to have a larger proportion of fast transactions and also a larger proportion of very slow transactions.
Indeed,  it results that $43\%$ of the transactions are still not included in the Blockchain after 1h from the first time they were seen and, remarkably, $20\%$ of the transactions were still not included after 30 days, revealing therefore a great inefficiency in the system (this statistics is reported in Fig. \ref{cumulativeInclusion}).

A slightly different outcome is achieved if, instead of the number of transactions, we measure the fraction of transferred value that is included in the Blockchain after a given amount of time (Fig. \ref{cumulativeInclusionValue}).
In this case, we note that the process is still rather slow but most of the value is included in the Blockchain within $3h$ ($93\%$) and after $30$ days only $0.1\%$ of value is left to be included.
This apparent inconsistency with the previous results is caused by the fact that the tail of the probability distribution for long waiting transactions is mostly populated by transactions containing only very small amounts as indicated by Fig. \ref{BlocksValue}.

We verified that fees (computed as the difference between total value inserted in the transaction minus the total value paid) play a minor role on the time a transaction takes to be included in the Blockchain. This is reported in Fig. \ref{BlocksFee} where we observe that some transactions associated with high fees have very long waiting times and, {\it vice-versa}, transactions with small fees are processed rather  rapidly.

\section{Conclusions}
By monitoring the Bitcoin network activity during a period of one week and by following the dynamics of inclusions of transactions within the Blockchain during the following three months we unveiled strong inefficiencies. 
The Bitcoin system fails in taking accurate record of the transactions with some of them taking months before being recorded in the Blockchain.
{We note that this inaccurate recording does not seem to be caused by the fact that block size is limited to $1 MB$ and only few thousands transaction can be included into a block.
It seems indeed that the network is not saturated yet, with average block size  $0.8 MB$, with only 3\% of blocks exceeding $0.99 MB$ band and even with some blocks without transactions \cite{blockchain.info}.} 
The transactions most affected by the delay are those of small value, even if we observed  some large transactions that have being recorded with delays of over one month. 
This poses serious questions not only on the efficiency of the Bitcoin system in itself but also on the possibility of using Bitcoin  as a reliable time-stamping system  where small value transactions are used to register operations outside the Bitcoin system. 

We conclude that such inefficiency in the Bitcoin system is most likely due to lack of sufficient incentives for peers and miners to verify, propagate and record transactions.  
Indeed, peer contribution to the Bitcoin network is mostly driven by mining rewards and efficient record keeping is not incentivised.

\bibliographystyle{IEEEtran}

\end{document}